\documentclass[reprint,aps,prd,twocolumn,groupedaddress,preprintnumbers,amsmath,amssymb]{revtex4-2}

\usepackage{graphicx}
\usepackage{dcolumn}
\usepackage{bm}
\usepackage{hyperref}
\usepackage{cleveref}
\usepackage{xcolor}
\usepackage{physics}
\usepackage{multirow}
\usepackage{subcaption}
\usepackage{float} 
\usepackage{orcidlink}

\newcommand{\Var}{\ensuremath{\mathrm{Var}}}
\newcommand{\SU}{\mathrm{SU}}
\newcommand{\CP}{\mathrm{CP}}
\newcommand{\nstep}{n_{\mathrm{step}}}

\newcommand{\tauint}{\tau_{\mathrm{int}}}

\newcommand{\DKL}{\Tilde{D}_{\mathrm{KL}}}
\newcommand{\ESS}{\mathrm{ESS}}

\newcommand{\Pf}{\mathcal{P}_{\mathrm{f}}}
\newcommand{\Pre}{\mathcal{P}_{\mathrm{r}}}
\newcommand{\Wd}{W_{\mathrm{d}}}
\newcommand{\U}{\mathcal{U}}

\begin{document}

\title{Scaling of Stochastic Normalizing Flows in $\SU(3)$ lattice gauge theory}


\author{Andrea Bulgarelli\orcidlink{0009-0002-2917-6125}}
\affiliation{Dipartimento di Fisica,  Universit\`a degli Studi di Torino and INFN, Sezione di Torino, Turin, Italy}

\author{Elia Cellini\orcidlink{0000-0002-5664-9752}}
\affiliation{Dipartimento di Fisica,  Universit\`a degli Studi di Torino and INFN, Sezione di Torino, Turin, Italy}

\author{Alessandro Nada\orcidlink{0000-0002-1766-5186}}
\affiliation{Dipartimento di Fisica,  Universit\`a degli Studi di Torino and INFN, Sezione di Torino, Turin, Italy}

\date{\today}

\begin{abstract}
\noindent
Non-equilibrium Markov Chain Monte Carlo (NE-MCMC) simulations provide a well-understood framework based on Jarzynski's equality to sample from a target probability distribution. By driving a base probability distribution out of equilibrium, observables are computed without the need to thermalize. If the base distribution is characterized by mild autocorrelations, this approach provides a way to mitigate critical slowing down. Out-of-equilibrium evolutions share the same framework of flow-based approaches and they can be naturally combined into a novel architecture called Stochastic Normalizing Flows (SNFs). In this work we present the first implementation of SNFs for $\SU(3)$ lattice gauge theory in 4 dimensions, defined by introducing gauge-equivariant layers between out-of-equilibrium Monte Carlo updates. The core of our analysis is focused on the promising scaling properties of this architecture with the degrees of freedom of the system, which are directly inherited from NE-MCMC. Finally, we discuss how systematic improvements of this approach can realistically lead to a general and yet efficient sampling strategy at fine lattice spacings for observables affected by long autocorrelation times.
\end{abstract}


\maketitle

\section{Introduction and motivation}
\label{sec:intro}

In the last decades numerical Monte Carlo simulations of lattice-regularized quantum field theories have established themselves as the main non-perturbative tool to study strongly-interacting theories from first principles. In the context of Quantum Chromodynamics (QCD) on the lattice, Markov Chain Monte Carlo simulations have provided a robust framework for high-precision theoretical predictions that are routinely compared to experimental results, see for example refs.~\cite{FlavourLatticeAveragingGroupFLAG:2021npn, FlavourLatticeAveragingGroupFLAG:2024oxs}. 
However, despite the great success, lattice MCMC calculations still suffer from considerable shortcomings: for example, when simulating the theory close to a critical point the \textit{autocorrelation} of a sequence of field configurations sampled via MCMC diverges, giving rise to a phenomenon usually denoted as critical slowing down~\cite{Wolff:1989wq}. In order to quantitatively state this issue, one introduces the integrated autocorrelation time $\tauint(\mathcal{O})$, which estimates the effectively independent samples on which the observable of interest $\mathcal{O}$ is measured. 
When approaching criticality, $\tauint(\mathcal{O})$ diverges with the correlation length of the system: in the particular case of a lattice theory simulated at a spacing $a$, one observes that 
$$\tauint(\mathcal{O}) \sim a^{-z}.$$
The value of the exponent $z$ depends on the MCMC algorithm and on the observable $\mathcal{O}$: if it is large enough, the cost of generating independent samples can grow in an unsustainable way when moving towards the continuum limit. 

The most glaring example of critical slowing down in lattice gauge theories is the so-called topological freezing~\cite{Alles:1996vn, DelDebbio:2004xh, Schaefer:2010hu}: when approaching the continuum, standard MCMC algorithm struggle to overcome the barriers that separate topological sectors and thus subsequent configurations are effectively stuck on the same value of the topological charge. This behavior manifests itself with extremely long autocorrelation times that scale with a large $z$ coefficient, or even exponentially.
Open Boundary Conditions~\cite{Luscher:2011kk, Luscher:2012av} successfully mitigate the freezing of the topological charge by removing the barriers between different topological sectors; however, they possess their own limitations (requiring a large temporal extent and an analysis of boundary effects) and they are also aimed uniquely at the sampling of topological quantities; for other approaches aimed at topological freezing see for example refs.~\cite{Laio:2015era, Eichhorn:2023uge, Albandea:2021lvl}. A fully satisfying general framework to tackle critical slowing down in lattice field theories simulations is thus still missing.

Flow-based samplers~\cite{Cranmer:2023xbe} have recently emerged as a promising and very general approach to simulate lattice field theories. A flow-based approach essentially consists in the construction of a map between a base distribution, which is either tractable or characterized by mild autocorrelations, and a target distribution that is in turn affected by critical slowing down. If such a map can be designed with reasonable computational efficiency, it will effectively generate uncorrelated (or mildly-correlated) samples from a given Boltzmann distribution.
While the idea itself goes back to the concept of trivializing maps~\cite{Luscher:2009eq}, recent advances in the construction and training of deep neural networks enabled the search for much more general and complex maps.
In particular, in recent years it was shown that building on Normalizing Flows (NFs)~\cite{rezende2015variational}, a class of deep generative models, it was possible to design samplers able to naturally compute not just expectation values, but partition functions as well~\cite{Albergo:2019eim, Nicoli:2020njz, Nicoli:2023qsl}. 
In particular, in the last years the effectiveness of NFs has been widely investigated using different approaches on a number of lattice models, ranging from scalar theories~\cite{Albergo:2019eim, Nicoli:2020njz, DelDebbio:2021qwf, Gerdes:2022eve, Singha:2022icw, Chen:2022ytr, Caselle:2023mvh, Albandea:2023wgd, Bulgarelli:2024yrz}, gauge theories~\cite{Kanwar:2020xzo, Boyda:2020hsi, Bacchio:2022, Singha:2023xxq, Abbott:2023thq, Gerdes:2024rjk} also including fermions~\cite{Albergo:2021bna, Finkenrath:2022ogg, Albergo:2022qfi} and QCD~\cite{Abbott:2022zhs, Abbott:2024kfc}. 
NFs are also widely popular in quantum chemistry, where they are generally referred to as Boltzmann Generators~\cite{noe:2019blt} and are used to address challenges similar to those in lattice field theory, such as sampling complex energy landscapes~\cite{Invernizzi_2022} and computing free energies~\cite{Wirnsberger_2020}.
However, despite the benefits demonstrated on lower dimensional theories and the potential to address critical slowing down in a general way, flow-based samplers currently suffer from poor scaling when applied to large lattice volumes, see refs.~\cite{DelDebbio:2021qwf,Abbott:2022zsh, Komijani:2023fzy, Abbott:2023thq}.

An alternative approach to flow-based methods is represented by non-equilibrium calculations based on the exact equality discovered by C.~Jarzynski more than two decades ago~\cite{Jarzynski1997, Jarzynski1997_2}. It relates the difference in free energy between two thermodynamic states with the exponential average of the work performed on the system when driving it from one state to the other and it represents a true non-equilibrium result, as during such evolutions the system is not in thermodynamic equilibrium. In ref.~\cite{Caselle:2016wsw} out-of-equilibrium Monte Carlo simulations were applied for the first time in lattice field theory to compute the interface free energy in the $\mathbb{Z}_2$ gauge theory. Over the last decade these methods have been employed with great effectiveness for high-precision lattice calculations across a variety of domains, including the equation of state of $\SU(3)$ Yang-Mills theory~\cite{Caselle:2018kap}, the running coupling in $\SU(N)$ gauge theories~\cite{Francesconi:2020fgi} and the entanglement entropy in gauge theories~\cite{Bulgarelli:2023ofi, Bulgarelli:2024onj}.
More recently, such non-equilibrium evolutions have been exploited also to generate samples arbitrarily close to the target distribution via a reweighting-like mechanism. 
This provides a novel framework to mitigate critical slowing down by driving configurations from tractable distributions towards distributions that are affected by large autocorrelations at equilibrium. 
While sharing several similarities with flow-based approaches, the mapping in this case is purely stochastic and relies on results from non-equilibrium statistical mechanics~\cite{Jarzynski1997, Crooks_1999}.
This approach, that in the following we denote as Non-Equilibrium Markov Chain Monte Carlo (NE-MCMC), has recently been applied to mitigate topological freezing in $\CP^{N-1}$ models in two dimensions~\cite{Bonanno:2024udh} using out-of-equilibrium evolutions in the boundary conditions; the same approach is currently being extended to $\SU(3)$ pure gauge theory~\cite{Bonanno:2024fkn}.

Among the many advances related to the use of non-equilibrium processes based on Jarzynski's equality, a huge effort in the context of stochastic thermodynamics has been devoted to the development of optimal protocols that minimize the work spent during the evolution. This issue has been studied with several approaches~\cite{Schmiedl_2007, Gomez_Marin_2008, Aurell_2011, Aurell_2012, Bonanca_2018, Kamizaki_2022} and in particular using a geometric description, see refs.~\cite{Sivak_2012, Zulkowski_2012, Rotskoff_2015, Brandner_2020, Blaber_2020}; we refer to~\cite{Blaber_2023} for a review of this approach.

A well-known algorithm equivalent to NE-MCMC that has found many applications in different fields is Annealed Importance Sampling (AIS), first introduced in ref.~\cite{Neal2001} (for a modern reworking of AIS see also Sequential Monte Carlo~\cite{Dai2022}). In the machine learning community it lies at the foundation of diffusion models~\cite{sohl2015deep, ho2020denoising}, a popular class of deep generative models that has seen a recent application in lattice field theory as well~\cite{wang2024diffusion, Zhu:2024kiu,Aarts:2024rsl}. For generalizations of AIS with flows see also refs.~\cite{arbel2021annealed,Matthews:2022sds}.
Moreover, we mention also recent developments that leverage out-of-equilibrium statistical mechanics to improve sampling from Boltzmann distributions~\cite{albergo2024nets}.
Finally, Parallel Tempering techniques present some similarities with NE-MCMC, with the crucial difference that the intermediate probability distributions are always sampled at equilibrium in parallel while allowing for regular swaps of field configurations between them. This approach has for example been used to mitigate topological freezing combining simulations with open and periodic boundary conditions~\cite{Hasenbusch:2017unr} and has recently been extended to pure gauge theories~\cite{Bonanno:2020hht, Bonanno:2022yjr, Bonanno:2023hhp, Bonanno:2024nba} and QCD~\cite{Bonanno:2024zyn}; an extension of this approach involving Normalizing Flows has been presented in refs.~\cite{Invernizzi_2022, Abbott:2024mix}.

NE-MCMC simulations have demonstrated a clear scaling with the number of degrees of freedom involved in the transformation: in particular, in order to keep the variance of the NE-MCMC estimator of the observable fixed, the length of the out-of-equilibrium evolution has to increase approximately linearly with the degrees of freedom, as pointed out already in ref.~\cite{Bonanno:2024udh}.
Thus, a natural strategy to overcome the scaling limitations of flow-based approaches is to combine them into a novel architecture called Stochastic Normalizing Flows (SNFs), first developed in ref.~\cite{wu2020stochastic}. Conversely, this hybrid approach can be seen as a way to systematically accelerate and augment non-equilibrium Monte Carlo simulations; among related work connecting non-equilibrium simulations and deterministic transformations we mention refs.~\cite{Vaikuntanathan_2011, Nilmeier_2011}.
A key aspect of SNFs is their theoretical foundation, which is deeply rooted in non-equilibrium thermodynamics, as pointed out in ref.~\cite{Caselle:2022acb}. The same work provided an implementation of SNFs in the $\phi^4$ scalar lattice field theory and showed how this new architecture integrates the scalability and robustness of NE-MCMC with the expressiveness and efficiency of Normalizing Flows.

In the context of lattice field theory, SNFs have shown considerable success in recent applications to numerical simulations of Effective String Theory~\cite{Caselle:2024ent}, a two-dimensional scalar model used to study confinement, and in the computation of entanglement entropy in quantum field theory~\cite{Bulgarelli:2024yrz}. Building on these achievements, the focus of this work is to develop a SNF specifically tailored for the $\SU(3)$ lattice gauge theory in $d=3+1$ dimensions. Our main goal is to provide proof for a clear scaling of the costs underlying the sampling performed with NE-MCMC and SNFs, in particular in terms of the degrees of freedom involved in the non-equilibrium transformations.
This exploration aims to extend the utility of SNFs to more complex and higher-dimensional gauge theories, opening new avenues for their application in fundamental physics.
We start this manuscript by outlining the framework of NE-MCMC in section~\ref{sec:nemcmc} and we proceed with a detailed description of the SNF architecture developed for gauge theories in section~\ref{sec:snf}. In section~\ref{sec:results} we present an in-depth discussion of the numerical results obtained in the four-dimensional $\SU(3)$ lattice gauge theory for evolutions in which the lattice spacing is changed, with a particular focus on the scaling with the volume of the lattice. Finally, in section~\ref{sec:conclusions} we indicate the most promising directions in which this approach can be systematically improved and present feasible implementations to mitigate severe critical slowing down in state-of-the-art simulations.

\section{Non-equilibrium Markov Chain Monte Carlo}
\label{sec:nemcmc}

A non-equilibrium Markov Chain Monte Carlo (NE-MCMC) simulation is composed of an ensemble of out-of-equilibrium evolutions from a prior distribution $q_0 = \exp(-S_0)/Z_0$ to the target distribution $p=\exp(-S)/Z$ that we wish to sample from. Each of these evolutions is generated by drawing a sequence of $\nstep$ configurations $U_n$ using a varying transition probability. 
The starting configuration $U_0$ of this sequence is sampled from the prior $q_0$ and then driven towards the target distribution $p$ through a sequence of Monte Carlo updates defined by the transition probabilities $P_{c(n)}$. These transition probabilities depend on a set of parameters $c(n)$, known as the \textit{protocol}, which controls the intermediate steps of the system’s evolution, and drive the configurations sequence from the initial state toward the target distribution, i.e.,
\begin{equation}
\label{eq:NE-MCMC_seq}
  U_0 \stackrel{P_{c(1)}}{\longrightarrow} \; U_1 \;
  \stackrel{P_{c(2)}}{\longrightarrow} \; U_2 \;
  \stackrel{P_{c(3)}}{\longrightarrow} \; \dots \;
  \stackrel{P_{c(\nstep)}}{\longrightarrow} \; U_{\nstep} \equiv U
\end{equation}
where we note that the last transition probability is fixed to be the one of the target distribution, i.e., $P_{c(\nstep)} \equiv P_p$.

Formally, after fixing $\nstep$ and the protocol $c(n)$, we can introduce the forward $\Pf$ probability distribution of each evolution $[U_0,U_1,\dots,U]\equiv \U$, which can be written as
\begin{equation}
    \Pf[ \U] = \prod_{n=1}^{\nstep} P_{c(n)} (U_{n-1} \to U_n) \, ;
\end{equation}
similarly, the reverse probability distribution $\Pre$ takes the form of
\begin{equation} 
    \Pre[\U] = \prod_{n=1}^{\nstep} P_{c(n)} (U_{n} \to U_{n-1}).
\end{equation}
Provided that the transition probabilities $P_{c(n)}$ satisfy detailed balance, using these definitions we can state Crooks’ theorem~\cite{Crooks:1998,Crooks_1999}, which relates the probability distributions of forward and reverse evolutions to the dissipation of the evolution $\U$:
\begin{equation}
\label{eq:crooks}
\frac{q_0(U_0) \Pf[\U]}{p(U) \Pre [\U]} = \exp (W(\U) - \Delta F).
\end{equation}
In this expression, $\Delta F=-\log Z/Z_0$ is the dimensionless free energy difference\footnote{Across the manuscript $\Delta F$ is more precisely the free energy in units of the temperature: we remark that in a non-zero temperature setup this quantity is indeed connected to the actual free energy of the theory. We refer to ref.\cite{Caselle:2018kap} for an in-depth description on how to compute with NE-MCMC the pressure and the equation of state in the $\SU(3)$ pure gauge theory.} between the states described by $q_0$ and $p$, and $W$ represents the dimensionless work done on the system during its transformation from the initial to the final state:
\begin{equation}
\label{eq:work}
W(\U) = \sum_{n=0}^{\nstep-1} \left\{ S_{c(n+1)}\left[U_{n}\right] - S_{c(n)}\left[U_n\right] \right\}.
\end{equation}
We also introduce the dissipated work $\Wd(\U)=W(\U) - \Delta F$, which serves as a measure of the dissipation of the transformation between $q_0$ and $p$ uniquely defined by $\nstep$ and the protocol $c(n)$.
Interestingly, we can rewrite the work as
\begin{equation}
\label{eq:work2}
\begin{split}
W(\U) &= S(U)-S_0(U_0)-Q(\U),
\end{split}
\end{equation}
where $Q$ denotes the pseudo-heat exchanged during the evolution, defined as:
\begin{equation}
\label{eq:heat}
\begin{split}
Q(\U)&=\sum_{n=1}^{\nstep} \left\{ S_{c(n)}\left[U_{n}\right] - S_{c(n)}\left[U_{n-1}\right] \right\}.
\end{split}
\end{equation}
The quantity $Q$ tracks the energy difference at each step $n$ after updating the system with the transition probability defined by the protocol $c(n)$. It is the natural quantity that arises when deriving eq.~\eqref{eq:crooks} from the definition of forward and reverse probabilities. We remark that in order to derive eq.~\eqref{eq:crooks} the detailed balance condition must be satisfied by the Monte Carlo updates:
\begin{equation}\label{eq:detbal}
    \frac{ P_{c(n)} (U_{n-1} \to U_n) }{P_{c(n)} (U_{n} \to U_n-1)}=\frac{p_{c(n)}(U_n)}{p_{c(n)}(U_{n-1})}
\end{equation}
here $p_{c(n)} \propto \exp(-S_{c(n)})$ is the distribution fixed in the Markov kernels. 

Let us now focus on how to use NE-MCMC to compute the expectation value of an observable $\mathcal{O}$ over the probability distribution $p$. 
First, we introduce the average $\langle \dots \rangle_{\mathrm{f}}$ over all the possible forward evolutions drawn accordingly to $q_0 \Pf$, that we naturally define as:
\begin{equation} 
\begin{split}
  \langle \dots \rangle_{\mathrm{f}} &= \int \dd U_0 \dots \dd U \; q_0(U_0) \Pf[U_0,\dots,U] \; \dots \\
  &=  \int \dd \U \; q_0(U_0) \Pf[\U] \; \dots \\
\end{split}
\end{equation}
where we used the shorthand $\dd \U = \dd U_0 \dots \dd U$ for the integration over all possible configurations. 
Since the reverse sequence $U,\dots, U_0$ starts from configurations drawn accordingly to the target $p$, computing $\mathcal{O}(U)$ over $p$ or $\Pre$ is equivalent:
\begin{equation}
\begin{split}
    \langle \mathcal{O} \rangle_p 
    &=\int \dd U  \; p(U) \mathcal{O}(U) \\
    &=\int \dd \U \; p(U) \Pre[U, \dots, U_0] \;\mathcal{O}(U) \\
    &=\int \dd \U \; q_0(U_0) \Pf[\U] e^{-\Wd (\U)} \mathcal{O}(\U) \; ;
    \end{split}
\end{equation}
in the last step we used Crooks' theorem~\eqref{eq:crooks}.
We finally obtain the reweighting-like formula 
\begin{equation}
\label{eq:estimator}
    \langle \mathcal{O}(U) \rangle_p  =  \langle \mathcal{O}(\U) e^{-\Wd (\U)} \rangle_{\mathrm{f}}
\end{equation}
that tells us that how to correct for the probability distribution generated at the end of the evolution not corresponding exactly to $p$.
In order to determine $\Delta F$ we just set $\mathcal{O} = 1$, obtaining
\begin{equation}
1 = \langle e^{-\Wd(\U)} \rangle_{\mathrm{f}} 
\end{equation}
that in turn leads to Jarzynski's equality~\cite{Jarzynski1997}:
\begin{equation}
    \label{eq:jar}
    \langle e^{-W(\U)} \rangle_{\mathrm{f}} = e^{-\Delta F}.
\end{equation}
The computation of free energy differences between two distributions using Monte Carlo simulations is notoriously cumbersome, as it generally requires a large number of samples from intermediate distributions \textit{at equilibrium}. Thus, eq.~\eqref{eq:jar} represents a powerful estimator of free energy differences, as the intermediate distributions can be sampled without ever letting the system thermalize. 
In the context of lattice gauge theories Jarzynski's equality has seen a wide range of applications in the last few years, see refs.~\cite{Caselle:2016wsw, Caselle:2018kap, Francesconi:2020fgi, Bulgarelli:2023ofi, Bulgarelli:2024onj}.
Furthermore, eqs.~\eqref{eq:jar} and ~\eqref{eq:estimator} represent the main ingredients of NE-MCMC. i.e., they provide a framework to sample problematic distributions from non-equilibrium distributions in an unbiased way; we refer to ref.~\cite{Bonanno:2024udh} for a recent application on a model with frozen topology.

Note that no assumptions regarding thermalization were made during the construction of the NE-MCMC evolutions: the only exception is that if the prior distribution $q_0$ is sampled with a MCMC, it must be at equilibrium. While the starting and ending states of the transformations are fixed, there are in principle no restrictions on the protocol, either in the velocity of the evolution (i.e., $\nstep$) or in the functional form of $c(n)$; thus, the intermediate and final configurations $U_n$, with $0<n\leq \nstep$, are in principle allowed to lie arbitrarily far from equilibrium. 

In practice, however, the reliability of NE-MCMC estimators is deeply connected to how far from equilibrium evolutions defined by a given protocol are.
Namely, it is of the utmost importance to understand how the design of a NE-MCMC (in particular, different choices in terms of $\nstep$ and $c(n)$) can affect its performance.

An intriguing consequence of this non-equilibrium framework is that the average dissipated work $\Wd$ over all the possible evolutions of the NE-MCMC, is equivalent to the (reverse) Kullback-Leibler (KL) divergence between the probability densities of the forward and reverse evolutions, i.e.:
\begin{equation}
\label{eq:kl}
    \begin{split}
        \langle  \Wd (\U) \rangle_{\mathrm{f}} &= \langle \log \frac{q_0(U_0) \Pf(\U)}{p(U) \Pre(\U)} \rangle_{\mathrm{f}} \\
        & =\DKL(q_0(U_0) \Pf \| p(U) \Pre) \geq 0
    \end{split}
\end{equation}
It measures the similarity between two distributions and, in this non-equilibrium context, provides a way to study in a quantitative way the reversibility of the stochastic transformation we are employing.
Large dissipation, which thermodynamically corresponds to strong irreversibility, leads to a large KL divergence; conversely, when the evolution is fully reversible forward and reverse evolutions are indistinguishable, resulting in $\DKL(q_0 \Pf \| p\Pre)=0$. 
Intriguingly, since the KL divergence is positive-definite eq.~\eqref{eq:kl} implies the Second Law of thermodynamics for these non-equilibrium transformations, i.e.,
\begin{equation}
    \langle W(\U) \rangle_\mathrm{f} \geq \Delta F
\end{equation}
providing a natural interpretation in terms of the thermodynamics of Markov Chain Monte Carlo.

The KL divergence of eq.~\eqref{eq:kl} represents a crucial metric to determine the reliability of NE-MCMC when sampling a finite number of evolutions. To understand why, we start by looking at the probability distribution of the last configuration $U\equiv U_{\nstep}$, i.e., the one generated by a given protocol at the end of the evolution. We can write it as:
\begin{equation}
        q(U) = \int \dd U_0 \dots \dd U_{\nstep-1} \; q_0 (U_0) \Pf (\U)
\end{equation}
and it is in general intractable unless we let the system thermalize. In order to quantitatively describe the overlap between $q$ and the target $p$, we would like to estimate the KL divergence
\begin{equation}
\label{eq:kl2}
    \DKL (q \| p) = \int \dd \U \; q (U) \log \left( \frac{q (U)}{p (U) } \right) =  \langle \log \frac{q}{p} \rangle_{\mathrm{f}}.
\end{equation}
It is easy to prove that the divergence of eq.~\eqref{eq:kl} represents an upper bound for $\DKL (q \| p)$:
\begin{equation}
\begin{split}
    \DKL (q_0 \Pf \| p \Pre) - \DKL (q \| p) &= \langle \log \frac{q_0 \Pf }{p \Pre } - \log \frac{q}{p} \rangle_{\mathrm{f}} \\
    &= \langle \log \frac{q_0 \Pf }{q \Pre } \rangle_{\mathrm{f}} \geq 0
\end{split}
\end{equation}
where the last inequality holds since $\DKL (q_0 \Pf \| q \Pre)$ is itself a KL divergence.
It is then clear that studying the reversibility of a NE-MCMC with $\DKL (q_0 \Pf \| p \Pre)$ provides a reliable metric to evaluate the quality of the sampling of the target distribution.
In particular, protocols far from equilibrium lead to estimators of eqs.~\eqref{eq:jar} and ~\eqref{eq:estimator} with poor overlap with $p$, whereas quasi-reversible evolutions are guaranteed to sample it efficiently.

One can understand this by observing that irreversible evolutions characterized by $W(\U) \gg \Delta F$ will have a vanishing weight in eqs.~\eqref{eq:estimator} and \eqref{eq:jar}: conversely, transformations with negative dissipation, which violate the Second Law ($W(\U) < \Delta F$), provide the largest weights; a good sampling of such evolutions is instrumental for a reliable estimation of the non-equilibrium averages. We refer to appendix~\eqref{sec:work_distro} for a more detailed discussion in this direction.

Another metric that we will use throughout this work is the Effective Sample Size (ESS), that in the context of NE-MCMC is defined as
\begin{equation}
\label{eq:ess}
    \hat\ESS = \frac{\langle \exp(-W) \rangle_{\mathrm{f}}^2}{\langle \exp(-2W)\rangle_{\mathrm{f}}} = \frac{1}{\langle \exp(-2\Wd) \rangle_{\mathrm{f}}}.
\end{equation}
This quantity is an approximate estimator of the ratio
$$ \frac{\Var(\mathcal{O})_p}{\Var(\mathcal{O})_{\mathrm{NE}}} = \ESS$$
where the variance at the numerator is the one of an observable $\mathcal{O}$ computed directly on the target distribution $p$, whereas at the denominator we find the variance of the estimator of the right-hand side of eq.~\eqref{eq:estimator}. 
The interpretation in terms of NE-MCMC evolutions is straightforward: it is indeed easy to see that
\begin{equation}
\label{eq:essvar}
    \Var(\exp(-W)) = \left( \frac{1}{\hat\ESS}-1 \right) \exp(-2\Delta F) \geq 0.
\end{equation}
The positivity of the variance directly implies $0\leq\hat\ESS \leq 1$. More generally, this quantity gives us a quantitative description of the distribution of the weights of $\exp(-W(\U))$ across different evolutions: indeed, investigating this distribution represents an effective way to monitor the overlap with the target distribution.

One simple and effective approach to either reduce the KL divergence of eq.~\eqref{eq:kl} or to increase the ESS of eq.~\eqref{eq:ess} is to increase the number of steps in the NE-MCMC protocols, i.e., going towards the $\nstep \to \infty$ limit: smoother, “quasi-static” transformations keep the thermodynamic transformations close to equilibrium, thereby maintaining vanishing dissipation. A more elaborate and potentially much more effective strategy is to optimize the functional form of the protocol $c(n)$ itself: this approach remains one of the main challenges for implementing more efficient NE-MCMC algorithms. 

Since NE-MCMC effectively builds maps between samples, the autocorrelation on samples obtained from the estimators of eqs.~\eqref{eq:estimator} and \eqref{eq:jar} cannot be larger that the one of the starting distribution $q_0$. 
In this sense this approach can be exploited in a novel strategy to mitigate critical slowing down: for instance, configurations at coarser lattice spacing can be sampled with reasonable autocorrelation and each of them can then be used as starting point for a NE-MCMC. Eq.~\eqref{eq:estimator} can be used to compute unbiased expectation values at at finer lattice spacing, where we expect critical slowing down to be more severe. 

We stress that the implementation of the protocol is not limited to changes in the coupling of the theory (controlling the lattice spacing); it can also involve variations in the geometry of the lattice configurations. 
In ref.~\cite{Caselle:2016wsw} the free energy of an interface was computed changing the boundary conditions of the lattice, while in refs.~\cite{Bulgarelli:2023ofi,Bulgarelli:2024onj}, the entanglement entropy of lattice quantum field theories was computed using evolutions that connect replica spaces with different geometries. Additionally, ref.~\cite{Bonanno:2024udh} addresses directly topological freezing by leveraging non-equilibrium evolutions between a system with open boundary conditions and one with periodic boundary conditions.

\section{Stochastic Normalizing Flows}
\label{sec:snf}

NE-MCMC evolutions may require a large number of Monte Carlo steps to provide a reliable sampling tool, especially in the case of non-optimal protocols: in this section, we outline a variational approach to augment NE-MCMC calculations with the aim of making them less computationally expensive in a systematic fashion. 
In particular, our proposal is to compose deterministic, parametric functions $g_n$, each depending on a set of parameters $\{ \rho^{(n)} \}$, with the Monte Carlo updates of the NE-MCMC. In practice, the sequence of eq.~\eqref{eq:NE-MCMC_seq} becomes 
\begin{equation}
   U_0 \stackrel{g_1}{\longrightarrow} \; g_1(U_0) \;
  \stackrel{P_{c(1)}}{\longrightarrow} \; U_1 \;
  \stackrel{g_2}{\longrightarrow} \; g_2(U_1) \;
  \stackrel{P_{c(2)}}{\longrightarrow} \; \dots \;
\end{equation} 

 The idea of combining stochastic and deterministic transformations to enhance NE-MCMC-like methods was already investigated in refs.~\cite{Vaikuntanathan_2011, Nilmeier_2011}. The challenges of this approach lie not only in the implementation but, most importantly, in the optimization of the functions $g_n$.
 Nevertheless, the rapid progression of deep learning has provided an elegant tool to design a variational augmentation of NE-MCMC in the form of Normalizing Flows (NFs)~\cite{rezende2015variational}. 
 NFs are deep generative models that can be used for density estimation~\cite{rezende2015variational} and to approximate Boltzmann distributions~\cite{Albergo:2019eim,Nicoli:2020njz} through the construction of a sequence of parametric diffeomorphisms $g_n$. 
 The key feature of NFs is that the change in density between the input $U_{n-1}$ and the output $g_n(U_{n-1})$ of the function can be exactly computed:
 \begin{equation}
 \label{eq:nf_jacobian}
     \frac{q_n(g_n(U_{n-1}))}{q_{n-1}(U_{n-1})}=J_{g_n}^{-1}(U_{n-1})
 \end{equation}
 where $J_{g_n}$ is the determinant of the Jacobian of $g_n$. 

In our approach Normalizing Flows (NFs) are combined with NE-MCMC to build Stochastic Normalizing Flows (SNFs)~\cite{wu2020stochastic, Caselle:2022acb} by interleaving the parametric diffeomorphisms with the Markov kernels of the out-of-equilibrium algorithms. 
It is possible to generalize Crooks' theorem~\eqref{eq:crooks} setting the transition probability for the $g_n$ to be
\begin{equation}
    P_{g_n}(U_{n-1}\to U_n)=\delta(U_n-g_n(U_{n-1}))
\end{equation}
and modifying $\Pf$ and $\Pre$ appropriately.
Eq.~\eqref{eq:crooks} then still holds if we naturally generalize the work $W$ of eq.~\eqref{eq:work2} as:
\begin{equation}
\label{eq:work_snf}
\begin{split}
    W(\U)= S(U) - S_0(U_0) - Q(\U) -\log J(\U),
    \end{split}
\end{equation}
where the new term:
\begin{equation}
    \log J(\U) = \sum_{n=1}^{\nstep} \log J_{g_n}(U_{n-1})
\end{equation}
corresponds to the sum of the logarithms of the Jacobian determinant of the NF layers and comes from the use of the change of variables theorem and from eq.~\eqref{eq:nf_jacobian}.
More explicitly we have  that
\begin{equation}
\label{eq:work2_snf}
\begin{split}
 W(\U) =  \sum_{n=0}^{\nstep-1} &S_{c(n+1)}\left[g_{n+1} (U_n) \right] - S_{c(n)}\left[U_n \right] + \\
 & -\log J_{g_{n+1}}(U_n).
\end{split}
\end{equation}

Setting $g_n$ to the identity we immediately recover NE-MCMC. Naturally, we want to tune the parameters $\{ \rho^{(n)} \}$ of the functions $g_n$ to improve the performances of NE-MCMC.
In a very natural way, we optimize the SNF by computing the appropriate generalization of the KL divergence of eq.~\eqref{eq:kl} and minimizing it:
 \begin{equation}
      \textrm{min}_{ \{ \rho \} } \DKL( q_0 \Pf \| p \Pre) = \textrm{min}_{\{ \rho \}} \langle \Wd (\U ) \rangle_\mathrm{f} 
 \end{equation}
Numerically, this optimization tunes the parameters of the coupling layers in such a way to minimize the dissipated work $\Wd$ of each trajectory as much as possible. From a thermodynamic perspective, the optimized SNF generates a more reversible set of non-equilibrium evolutions compared to a NE-MCMC with the same protocol. 
SNFs thus offer a way to optimize NE-MCMC, forcing the evolutions to be as reversible as possible through deterministic maps, while potentially inheriting the scalability of the stochastic updates. 

A similar approach to SNFs is the Continual Repeated Annealed Transport Flow (CRAFT)~\cite{Matthews:2022sds} method, in which Sequential Monte Carlo is combined with NFs. 
In this approach, each NF layer of the variational NE-MCMC is trained separately by minimizing the KL divergence $\DKL(q_n| p_{c(n)})$, where $U_n=g_n(U_{n-1})$ and $U_{n-1}\sim p_{c(n-1)}$, to optimize the transition between the preceding and successive Markov kernel of the deterministic transformation. 

In lattice field theory, SNFs have been first applied to the $\phi^4$ scalar field theory in two dimensions~\cite{Caselle:2022acb}, and then to more challenging theoretical setups such as the sampling of Effective String Theories on the lattice~\cite{Caselle:2024ent} and the computation of entanglement entropy in scalar field theories in two and three dimensions~\cite{Bulgarelli:2024yrz}.
We refer to the latter for an in-depth comparison between NE-MCMC, NFs and SNFs in a setting where the number of relevant degrees of freedom is still relatively low.

\subsection{Gauge Equivariant Layers}

One of the most popular ways to implement the diffeomorphisms $g_n$ is using the so-called coupling layers~\cite{dinh2015, Dinh:2017}, a class of transformations that, by leveraging certain masking patterns, provides both invertibility and tractable Jacobian determinants and can be designed to encode specific symmetries into the transformation. Generally, encoding symmetries directly into machine learning models can enhance training efficiency and improve model quality, rather than having the model gradually learn these symmetries during training~\cite{Taco:2016equiv}. In flow-based sampling, symmetries can be incorporated into the architectures by making the NF equivariant~\cite{kohler2019equivariant, Kanwar:2020xzo, Abbott:2023thq}; in other words, the diffeomorphism $g_n$ commutes with the chosen symmetry.

In $\SU(3)$ gauge theory, link variables $U_{\mu}(x)$ transform under gauge transformations as
\begin{align}
U_{\mu}(x) \to \Tilde{U}_{\mu}(x)= G_{\mu}(x) U_{\mu}(x) G^{\dagger}_{\mu}(x+\hat{\mu})
\end{align}
where $G_{\mu}(x)$ is a generic $\SU(3)$ matrix. An equivariant layer $g_n$ thus satisfies:
\begin{align}
g_n(\Tilde{U}_{\mu}(x)) = G_{\mu}(x) g_n(U_{\mu}(x)) G^{\dagger}_{\mu}(x+\hat{\mu}),
\end{align}
and, if the input distribution is gauge invariant, the output density $q_n$ remains invariant with respect to the gauge symmetry, such that $q_n(U_{\mu}(x))=q_n(\Tilde{U}_{\mu}(x))$. One way to implement invertible gauge equivariant layers is to construct coupling layers that leverage transformations inspired by the stout smearing procedure~\cite{Morningstar:2003gk}. Stout smearing has been proposed as the base of residual convolutional neural networks for lattice gauge theory~\cite{he2015deep,Tomiya:2021ywc}, and, with the appropriate masking patterns, they can be exploited as gauge-equivariant coupling layers~\cite{Abbott:2023thq}.
In this work, we provide the first implementation of a SNF for the $\SU(3)$ theory in 4 dimensions using this approach, which we will examine in detail in the rest of this section.
We remark however that smearing-like transformations are not the only equivariant architecture studied in the context of NFs for lattice gauge theory and an alternative approach is the spectral flow, introduced in ref.~\cite{Kanwar:2020xzo,Boyda:2020hsi}. 
Furthermore, instead of using a sequence of discrete transformations, an elegant approach is through the use of Continuous Normalizing Flows (CNFs), in which the diffeomorphism $g$ is the solution of a ODE: in particular, the architecture proposed in ref.~\cite{Bacchio:2022} can be seen as a way of constructing more general Wilson flows~\cite{Luscher:2009eq}, and in ref.~\cite{Gerdes:2024rjk} this approach is generalized even further. 

In our implementation of gauge-equivariant layers each transformation $g_n$ is composed of a sequence of masks, each of which defines a set of ``active'' links, that undergo the transformation, while the remaining ``frozen'' links are left untouched. In this case we made use of 8 masks, defined by an even-odd decomposition for each space-time direction. 
Active links are transformed according to a stout-smearing transformation~\cite{Morningstar:2003gk}
\begin{equation}
    U'_\mu (x) = g_n (U_\mu (x)) = \exp \left(i Q_\mu^{(n)} (x) \right) \, U_\mu (x),
    \label{eq:coupling_layer_stout_smearing_transformation}
\end{equation}
with
\begin{equation}
\begin{split}
     Q_\mu^{(n)} (x) =  \frac{i}{2} \left( (\Omega^{(n)}_\mu (x))^\dagger - \Omega^{(n)}_\mu (x) \right) + \\
     - \frac{i}{2N} \Tr \left( (\Omega^{(n)}_\mu (x))^\dagger - \Omega^{(n)}_\mu (x) \right) 
\end{split}
\end{equation}
being Hermitian and traceless, while $\Omega_\mu^{(n)} (x)$ is a sum of untraced loops based on $x$. 
We have that
\begin{equation}
    \Omega_\mu^{(n)} (x) = C_\mu^{(n)} (x) U^\dagger_\mu (x)
\end{equation}
and $C_\mu^{(n)} (x)$ is a weighted sum over staples composed exclusively of frozen links:
\begin{equation}
\begin{split}
\label{eq:cmu_stout}
    C_\mu^{(n)} (x) =  \sum_{\nu \neq \mu} \rho^{(n)}_{\mu \nu}(x)  \left(  U_\nu(x) U_\mu(x+\hat{\nu}) U_\nu^\dagger(x + \hat{\mu}) \right. \\
    \left. + U^\dagger_\nu(x-\hat{\nu})U_\mu(x-\hat{\nu})U_\nu(x-\hat{\nu}+\hat{\mu}) \right).
\end{split}
\end{equation}
One possible way to generalize this layer is to use different sets of untraced loops, which however requires changing the masking pattern as well. 

The optimal stout-smearing parameters $\rho^{(n)}_{\mu \nu}(x)$ in eq.~\eqref{eq:cmu_stout} have to be determined using a training procedure. In this work we simply set them to be invariant under translations and (discrete) rotations, i.e., $\rho^{(n)}_{\mu \nu}(x)=\rho^{(n)}$, for a total of 8 parameters per layer.
This construction makes the Jacobian of the transformation easily tractable; for a discussion of invertibility we refer to the discussion on the link-level flow in ref.~\cite{Abbott:2023thq}.

Let us take a closer look at the Jacobian of the transformation of eq.~\eqref{eq:coupling_layer_stout_smearing_transformation}. The notation we are using here is different, though equivalent, to~\cite{Tomiya:2021ywc}. Explicitly introducing the color indices and neglecting the index $(n)$ of the layer, eq.~\eqref{eq:coupling_layer_stout_smearing_transformation} becomes (a sum over repeated indices is implicit)
\begin{align}
    [U'_\mu(x)]^a_b = [\exp \left(i Q_\mu(x) \right)]^a_i \, [U_\mu (x)]^i_b.
    \label{eq:coupling_layer_in_coordinates}
\end{align}
It is convenient to see the previous equation as a map between the coefficients instead of a map between matrices, as the two spaces are isomorphic. The Jacobian matrix is then defined as
\begin{align}
    J^{(\mu x a b)}_{(\nu y c d)} = \frac{\partial [U'_\mu(x)]^a_c}{\partial [U_\nu(y)]^b_d}.
\end{align}
Using~\eqref{eq:coupling_layer_in_coordinates} we get
\begin{align}
    J^{(\mu x a b)}_{(\nu y c d)} = \frac{\partial [\exp \left(i Q_\mu(x) \right)]^a_i}{\partial [U_\nu(y)]^b_d} [U_\mu(x)]^i_c + [\exp \left( iQ_\mu(x) \right)]^a_i \delta^i_b \delta^d_c.
\end{align}
The first term of the previous equation multiplying  $[U_\mu(x)]^i_c$ can be further expanded as
\begin{align}
    \frac{\partial [\exp \left(i Q_\mu(x) \right)]^a_i}{\partial [Q_\sigma(z)]^j_k} \, \frac{\partial [Q_\sigma(z)]^j_k}{\partial [\Omega_\tau(w)]^l_m} \left\{ \frac{\partial [\Omega_\tau(w)]^l_m}{\partial [U_\nu(y)]^b_d} - \frac{\partial [\Omega^\dagger_\tau(w)]^l_m}{\partial [U_\nu(y)]^b_d} \right\}.
\end{align}
As $\Omega$ is a linear function of the configurations, the term in curly brackets is easily computable, and the same holds for the derivative of $Q$ with respect to $\Omega$; $\partial \exp(iQ)/\partial Q$ is expressed in terms of the same matrices used to construct $\exp(iQ)$; for further details we point to refs.~\cite{Morningstar:2003gk, Tomiya:2021ywc}. Finally, the Jacobian of the transformation is easily obtained as the product of the Jacobian (with respect to the color indices) on every active link.

\section{Numerical results}
\label{sec:results}

We test the scaling of both NE-MCMC and SNFs architectures on simulations of the $\SU(3)$ Yang-Mills theory in 4 dimensions regularized on a lattice of physical volume $L^4$ and lattice spacing $a$, using the plaquette (Wilson) action. 
In all simulations the Monte Carlo update of choice is a composition of 1 step of heatbath plus 4 steps of over-relaxation; this is true both for NE-MCMC and SNFs. 
In this work we explore both approaches for transformations in the lattice spacing: the system is thermalized at a given initial value of the inverse bare coupling $\beta_0$ and then it is driven out of equilibrium using a linear protocol in $\beta$: 
\begin{equation}
  \beta(n) = \beta_0 + n \; \frac{\beta_t - \beta_0}{\nstep} \qquad n \in [1, \nstep],
\end{equation}
with $\beta_t>\beta_0$ being the target value of the inverse bare coupling.
We studied different transformations in the lattice spacing at different volumes: the ensembles are listed in table~\ref{tab:ensembles}. 
We remark that at these values of the inverse coupling, with the combination of heatbath and overrelaxation algorithms used in this work on a standard equilibrium simulation, topological observables are not frozen yet. We also point out that the same computational strategy was already used in ref.~\cite{Caselle:2018kap} for the computation of the pressure at non-zero temperature across the deconfinement transition.

\begin{table}[ht]
 \centering
 \begin{tabular}{|c|c|c|c|}
 \hline
  ensemble & $\beta$ & $a/r_0$ & $L/a$ \\
  \hline
  a2 & $5.896 \to 6.037$     & $0.225 \to 0.175$ & $10,12,16,20$ \\
  a3 & $6.02 \to 6.178$      & $0.180 \to 0.140$ & $10,12,16,20$ \\
  b0 & $6.0 \to 6.2$         & $0.186 \to 0.135$ & $12$ \\
  \hline
 \end{tabular}
 \caption{List of ensembles of transformations in the inverse coupling $\beta$ studied in this work, with the corresponding change in the lattice spacing $a$ in units of the Sommer scale $r_0$, using the scale setting from ref.~\cite{Necco:2001xg}. Volumes taken into consideration in this work are listed in the last column.}
 \label{tab:ensembles}
\end{table}

The `a2' and `a3' ensembles of transformations approximately correspond to a change in the physical size of the system of $1.8 \mathrm{fm} \to 1.4 \mathrm{fm}$ respectively for $L/a=16$ and $L/a=20$ lattices: on these transformations we studied the scaling of these architectures with the lattice volume $(L/a)^4$. 
The third ensemble of transformations (`b0') was used to give a preliminary analysis of the scaling of these samplers when moving in $\beta$ further away from the starting value.

\subsection{Training SNFs}

The search for the optimal parameters of a SNF architecture is performed through the minimization of the KL divergence of eq.~\eqref{eq:kl}. 
This minimization procedure has a clear thermodynamic interpretation: the parameters of the intermediate stout smearing layers are tuned in such a way that the evolution becomes as reversible as possible, i.e., towards $\langle \Wd(\U) \rangle_\mathrm{f} =0$.
It is interesting to note that the structure of the work of eq.~\eqref{eq:work2_snf} is rather peculiar, as it can be expressed as a sum over the steps $n$, each term containing a ``stochastic'' contribution from the difference of the action and the ``deterministic'' contribution from the Jacobian of the gauge-equivariant layer. 
It is then very natural to ask whether the training itself can be performed step by step, i.e., backpropagating the gradients not over the whole flow, but only between two MCMC updates: in practice, one uses the loss
$$ S_{c(n+1)}\left[g_{n+1} (U_n) \right] - S_{c(n)}\left[U_n \right] -\log J_{g_{n+1}}(U_n)$$
to train the parameters $\rho^{(n+1)}$ of the corresponding coupling layer $g_{n+1}$.
This is the strategy that we pursued in this work: in this way the memory usage needed to store the gradients during backpropagation is independent on $\nstep$ and depends uniquely on the depth of a single gauge-equivariant layer.
We note that this procedure is reminiscent of the one followed in the CRAFT architecture proposed in ref.~\cite{Matthews:2022sds}.

An important observation is that when training each layer separately as described above, the gradients used to update the parameters $\rho^{(n)}$ ignore the dependence of subsequent configurations on such parameters, with the resulting training of the whole architecture potentially being sub-optimal. 
Thus, we explicitly compared the results of our step-by-step training procedure with the standard one, in which gradients are propagated throughout the whole flow using the loss of eq.~\eqref{eq:work2_snf}. We found no discernible differences in the outcome of the two training procedures, suggesting that the optimal parameters of the coupling layers of SNFs depend mostly on the actions defining the ``neighboring'' MCMC updates that appear in each term in the sum of eq.~\eqref{eq:work2_snf}.
We also remark that backpropagating naively through Monte Carlo updates induces anyway a form of bias, as the accept/reject step is itself not differentiable.

\begin{figure}[t]
 \includegraphics[scale=0.55,keepaspectratio=true]{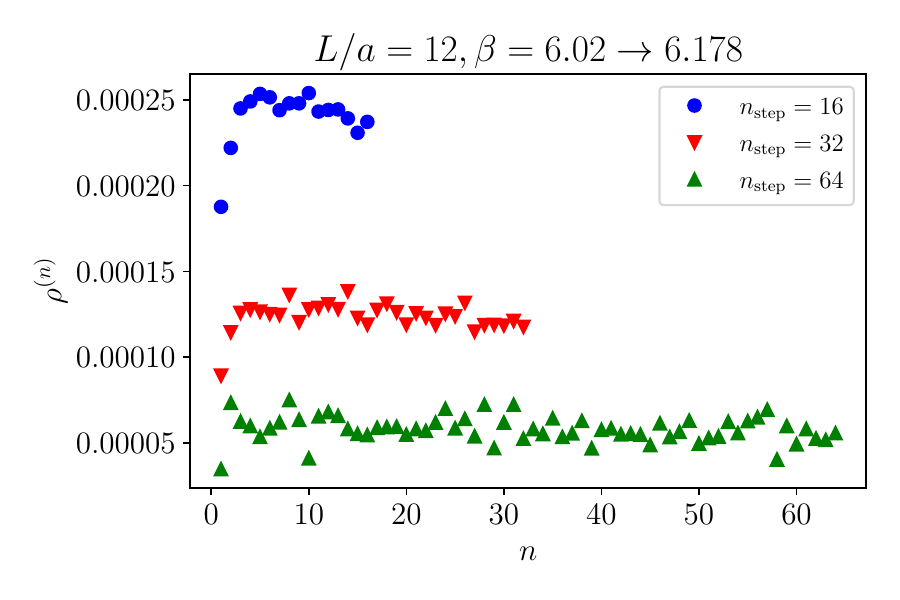}
 \includegraphics[scale=0.55,keepaspectratio=true]{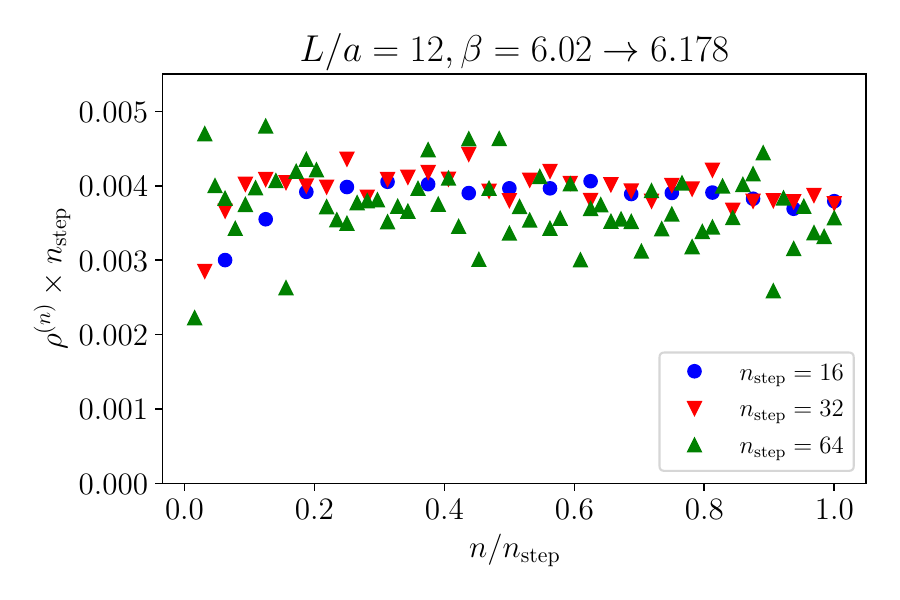}
 \caption{Value of the learned parameter $\rho^{(n)}$ along the SNF for the $n$-th layer (top panel) and the same value rescaled for $\nstep$ (bottom panel), for a training performed on a $L/a=12$ lattice for three values of $\nstep$ for a flow between $\beta=6.02$ and $\beta=6.178$.}
 \label{fig:rho_a3L12}
\end{figure}

In fig.~\ref{fig:rho_a3L12} we show results for the parameter $\rho^{(n)}$ of the stout smearing transformation of the $n$-th layer for architectures with $\nstep=16,32,64$ for $L/a=12$ and the `a3' ensemble. We recall that exactly one coupling layer is inserted between MCMC updates, so $n\in [1,\nstep]$. 
The length of the training was 800 epochs for $\nstep=16,32$ and 400 epochs for $\nstep=64$, with one epoch indicating one application of the Adam optimizer~\cite{Kingma:2014vow} to update the parameters; the learning rate was set to 0.0005.

Impressively, the values of $\rho^{(n)}$ roughly collapse on the same curve when rescaled with $\nstep$ and plotted against $n/\nstep$. The parameters obtained with $\nstep=64$ appear to be noisier, but the behavior for $\nstep=16,32$ is qualitatively very similar, with perhaps the only exception being the values in the first few layers. 
Due to the simplicity of this behavior, we implemented a global interpolation of these results for all $\nstep$ and used it to set the values of the smearing parameter on SNFs with larger $\nstep$. 
We tested the use of this interpolation for $\nstep>64$ with the results of a direct training procedure at the given value of $\nstep$, and we observed no differences in the relevant metrics. 
Thus, for the SNFs studied in this work we implemented the following transfer learning procedure: we trained only architectures with $\nstep=16,32,64$ for all ensembles and extrapolated the results for $\rho^{(n)}$ to larger $\nstep$. The behavior of the smearing parameters was qualitatively very similar also for the `a2' and `b0' ensembles of transformations in $\beta$.

While we performed this training procedure for each volume separately, we also noticed that transfer learning between the volumes seems to be possible as well. Namely, we obtained $\rho^{(n)}$ on $L/a=12$ lattices and then used the resulting parameters to sample the same transformation in $\beta$ but on a $L/a=20$: interestingly, the relevant metrics were compatible with an SNF with smearing parameters obtained on a training on a $L/a=20$ lattice.
For what concerning the `a2' and `a3' ensembles, the training of SNFs was performed only on the $L/a=20$ lattice and parameters (and their interpolation) were transferred to smaller volumes.

\subsection{Scaling of NE-MCMC and SNFs}

\begin{figure}[t]
 \includegraphics[scale=0.55,keepaspectratio=true]{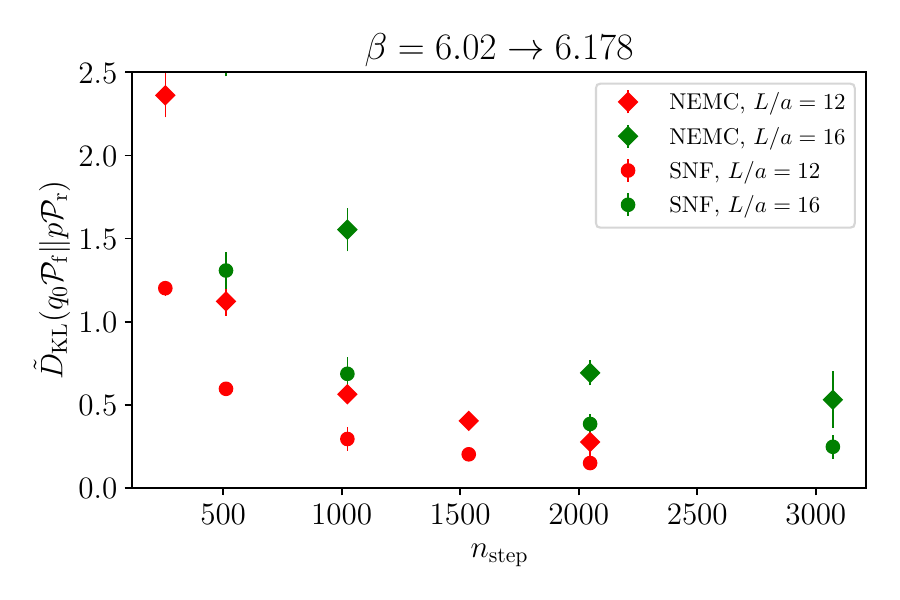}
 \includegraphics[scale=0.55,keepaspectratio=true]{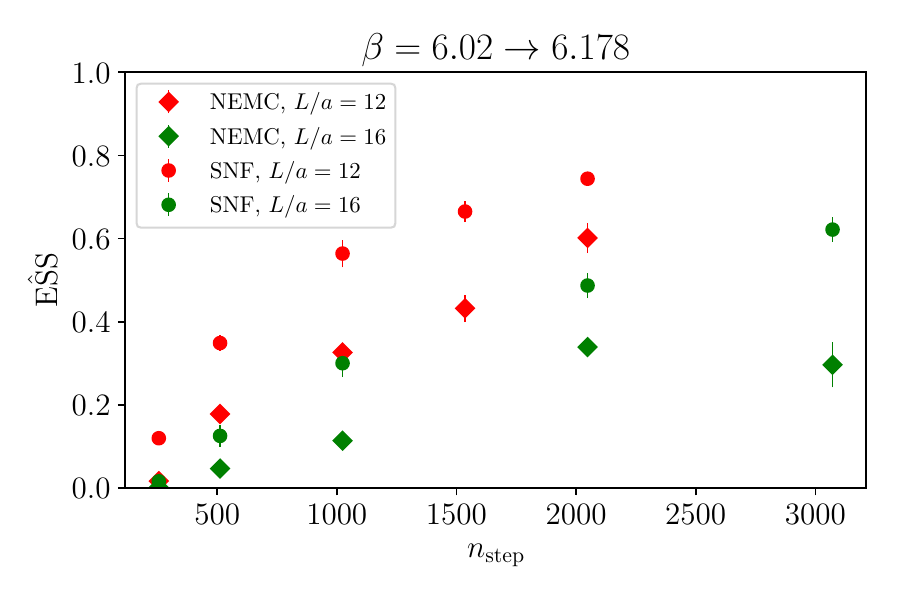}
 \caption{Results for the KL divergence (top panel) and for the ESS (bottom panel) for a flow between $\beta=6.02$ and $\beta=6.178$, for $L/a=12$ and $L/a=16$, for NE-MCMC (diamonds) and SNF (circles) architectures.}
 \label{fig:dkless}
\end{figure}

In this section we compare the performances of NE-MCMC and trained SNFs using the metrics introduced in section~\ref{sec:nemcmc}, for different ensembles of transformations in the lattice spacing.
In fig.~\ref{fig:dkless} results for the KL divergence of eq.~\eqref{eq:kl} and for the ESS of eq.~\eqref{eq:ess} are presented for two volumes and for several values of $\nstep$.
The dimensionless free energy $\Delta F$, needed to compute both quantities, is calculated separately for each flow using eq.~\eqref{eq:jar} using the appropriate definition of the dimensionless work $W$.

First of all, we observe that when $\nstep$ increases, i.e., when evolutions are performed closer to equilibrium, the KL divergence decreases for all architectures, reaching rather quickly values smaller than 1. As expected, the results indicate a convergence to $0$ as well, i.e., reversibility appears to be reached asymptotically. 
Furthermore, if we look at the metrics for each volume separately, it appears that SNFs are roughly a factor 2 more efficient than the standard NE-MCMC: indeed, the same value of $\DKL( q_0 \Pf \| p \Pre)$ is reached for about half the value of $\nstep$.

The same kind of improvement can be observed for the ESS: small values of $\nstep$ lead to $\hat\ESS<0.1$, which indicates a very poor sampling of the target distribution. As soon as $\nstep$ increases enough, depending on the volume that is tested, the same trend observed for the $\DKL$ appears. Both NE-MCMC and SNFs appear to approach the limit value $\hat\ESS=1$, with Stochastic Normalizing Flows being roughly twice as efficient as the purely stochastic counterpart.

\begin{figure}[t]
 \includegraphics[scale=0.55,keepaspectratio=true]{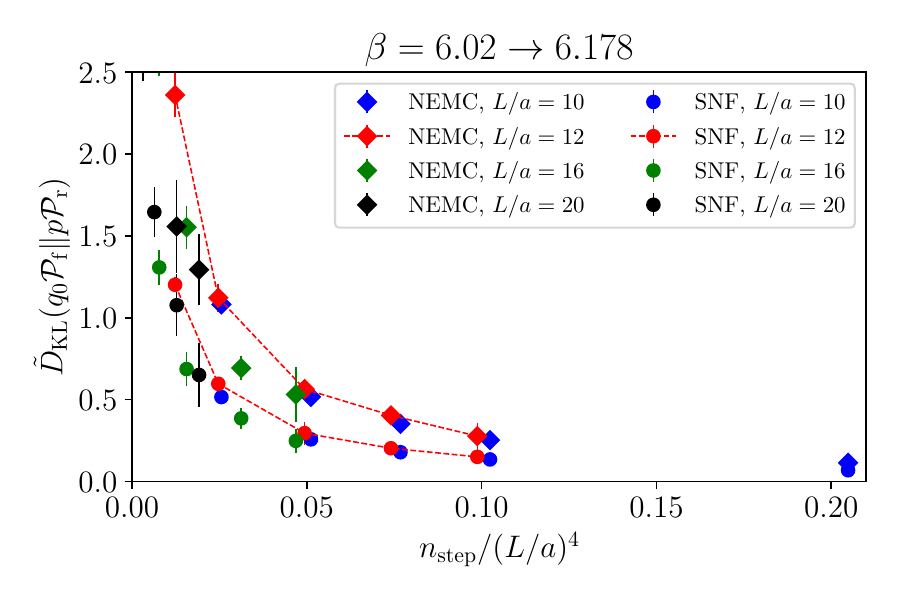}
 \includegraphics[scale=0.55,keepaspectratio=true]{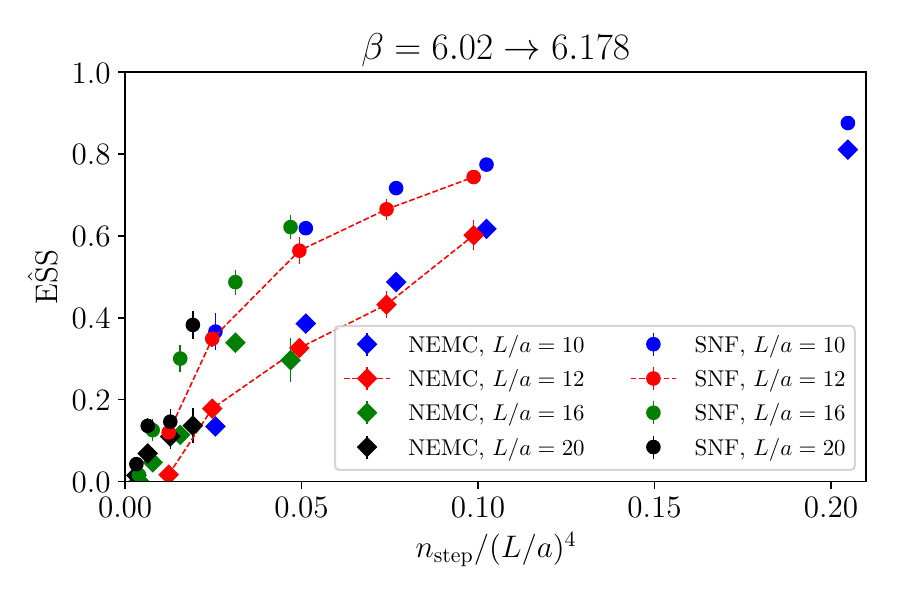}
 \caption{Results for the KL divergence (top panel) and for the ESS (bottom panel) for all the lattice sizes analyzed in this study for a flow between $\beta=6.02$ and $\beta=6.178$, plotted against $\nstep/(L/a)^4$, for NE-MCMC (diamonds) and SNF (circles) architectures. $L/a=12$ data are connected with a dashed line to guide the eye.}
 \label{fig:scaling_dkless}
\end{figure}

At this point it is important to stress two relevant details. First, that for the same value of $\nstep$ SNFs also perform a stout-smearing transformation of the links for each Monte Carlo update; in our experience the increase of the computational cost is however moderate, as in our implementation they are about 25\% as expensive as a full MCMC update.
Second, this improvement came at a very modest cost: as explained previously, the training was performed uniquely on values of $\nstep$ which are much smaller than the ones showed in fig.~\ref{fig:dkless}. In practical terms, the computational effort of the training is just a small fraction of the total effort used for the sampling for larger $\nstep$.
Before discussing the scaling of SNFs, it is natural to ask whether a NF (i.e., a flow without NE-MCMC steps) can reach similar performances. In this case, using the simple smearing transformation defined in eq.~\eqref{eq:coupling_layer_stout_smearing_transformation} it was not possible to create a satisfying NF, both by training the parameters from scratch or by using the ones obtained training a corresponding SNF.

Now we come to the analysis of the scaling. In fig.~\ref{fig:scaling_dkless} we plotted the results for the KL divergence and the ESS as a function of the number of steps normalized by the volume in lattice units. In this way, the results of fig.~\ref{fig:dkless} collapse on a unique curve, along those for $L/a=10$ and $L/a=20$: at least approximately, both metrics can be considered to be a function of $\nstep/(L/a)^4$.
In other words, if we keep the KL divergence or the $\hat\ESS$ fixed, i.e., we fix a target for our flow-based sampling, the number of steps $\nstep$ needed to reach this target scales with the volume of the lattice.
This behavior was already observed in ref.~\cite{Bonanno:2024udh}, where however the degrees of freedom changed throughout a non-equilibrium evolution are proportional to the size of a defect with open boundary conditions. In our case the inverse coupling $\beta$ is changed over all links, so the degrees of freedom of interest are proportional to the entire volume of the lattice. 
Crucially, the same scaling appears to hold also for trained SNFs: this is one of the main results of this work. We remark that the training was performed only for $\nstep=16,32,64$ and then the parameters were transferred to larger $\nstep$: yet, the same factor 2 of improvement over the NE-MCMC is always valid for all the volumes considered in this work: it appears that SNFs inherit the scaling with the volume (or, more precisely, with the degrees of freedom modified during an evolution) fully from the NE-MCMC. As it can be seen from fig.~\ref{fig:scaling_dkless_a2}, the same scaling holds also for transformations in $\beta$ at coarser lattice spacings.

\section{Conclusions and future prospects}
\label{sec:conclusions}

In this work we have presented the first implementation of Stochastic Normalizing Flows for the simulation of the $\SU(3)$ pure gauge theory in 4 spacetime dimensions, with a particular focus on their scaling properties with the degrees of freedom involved in the transformation between the base distribution and the target distribution that we want to sample from. 
This architecture was tested for the case of transformations in the lattice spacing of the theory, controlled by changing the inverse coupling $\beta$ of the theory over all the links on the lattice. The scaling relation that we found was that the metrics used to evaluate the quality of the sampling do not depend separately on the length of the transformation (i.e., $\nstep$) and the effective degrees of freedom (in this case proportional to the volume $(L/a)^4$), but just on the ratio $\nstep/(L/a)^4$.
This kind of scaling was already observed for NE-MCMC simulations in ref.~\cite{Bonanno:2024udh} and confirmed for the purely stochastic transformation also in this case: crucially, the same scaling is inherited also by SNFs, even if they involve deterministic transformations between the out-of-equilibrium Monte Carlo steps.

Another crucial aspect of SNFs is that they significantly improve the efficiency of NE-MCMC with a very small overhead cost spent on the training of the gauge-equivariant layers.
This was made possible by easily transferring the behavior of the parameters of the layers from fast, cheap transformations performed far from equilibrium to slower, more expensive ones with larger $\nstep$, without retraining them in the latter case.
These results points to an underlying structure in the smearing parameters $\rho^{(n)}$ which can be potentially understood without resorting to expensive and complicated training procedures.

More generally, this work paves the way for a systematic improvement of SNFs, while retaining the desirable scaling properties found in this analysis. We envision this improvement program to develop in two different directions: the first is towards a study of optimal protocols, which have been the subject of a vast literature in stochastic thermodynamics (see ref.~\cite{Blaber_2023} for a review) and for which several optimization methods already exist, see for example~\cite{Bonanca_2018,Kamizaki_2022}.
The second direction would be in the implementation of more complex and expressive gauge-equivariant layers: also in this case, recent advances in the development of Normalizing Flows for gauge theories provided generalizations of the smearing procedures in which the parameter $\rho^{(n)}$ itself is the output of a neural network~\cite{Abbott:2023thq}.
Although we leave the study of these improvements to future work, we wish to remark that a more comprehensive study of the relationship between gauge-equivariant coupling layers and the non-equilibrium Monte Carlo updates would be extremely valuable. An interesting aspect to examine in depth would be the performance of a SNF with a relatively small number of NE-MCMC steps but enhanced by more complex and expressive gauge-equivariant coupling layers. We expect that such an architecture would involve improved efficiency, as the flow would be less constrained by the protocol; conversely, it would likely require more significant training costs as well. Furthermore, it would be interesting to explore if (and when) the scaling with the degrees of freedom induced by the NE-MCMC breaks down, as the complexity of the network increases and/or the number of Monte Carlo steps decreases.
We point out that a comparison between NFs and SNFs has been carried out in~\cite{Bulgarelli:2024yrz}, albeit in a simpler setting.

Concerning practical applications to critical slowing down, we would like to point out that an efficient implementation of NE-MCMC for topological freezing has already been studied for transformations in the boundary conditions, both in $\CP^{(N-1)}$ models in 2 dimensions (see ref.~\cite{Bonanno:2024udh}) and in $\SU(3)$ in four dimensions (see preliminary results in ref.~\cite{Bonanno:2024fkn}).
This strategy involves a scaling of $\nstep$ with the size of the boundary which is switched from open to periodic boundary conditions, which in 4 dimensions amounts to a three-dimensional object, making it naturally cheaper than a transformation of $\beta$ over the whole lattice.
The generalization of the SNFs presented in this work for evolutions in the boundary conditions is relatively straightforward and will be the focus of future studies.

As a final word, we remark that the structural problem of critical slowing down is addressed in this work by essentially abandoning the assumption that a MCMC simulation has to be performed at equilibrium and designing a framework to exactly compute expectation values on non-thermalized samples. Furthermore, we envision this change of paradigm to go beyond the mitigation of diverging autocorrelations. In particular, this could lead to a (re)thinking of the thermalization process itself in MCMC simulations: such an analysis, obtained with controlled non-equilibrium processes, may play a role in master field simulations~\cite{Luscher:2017cjh, Giusti:2018cmp, Francis:2019muy, Bruno:2023vhs}, where expectation values are computed on few, very large lattices whose thermalization procedure requires particular care~\cite{Fritzsch:2021klm}.

\appendix

\section{Additional results and scaling in $\beta$}

\begin{figure}[t]
 \includegraphics[scale=0.55,keepaspectratio=true]{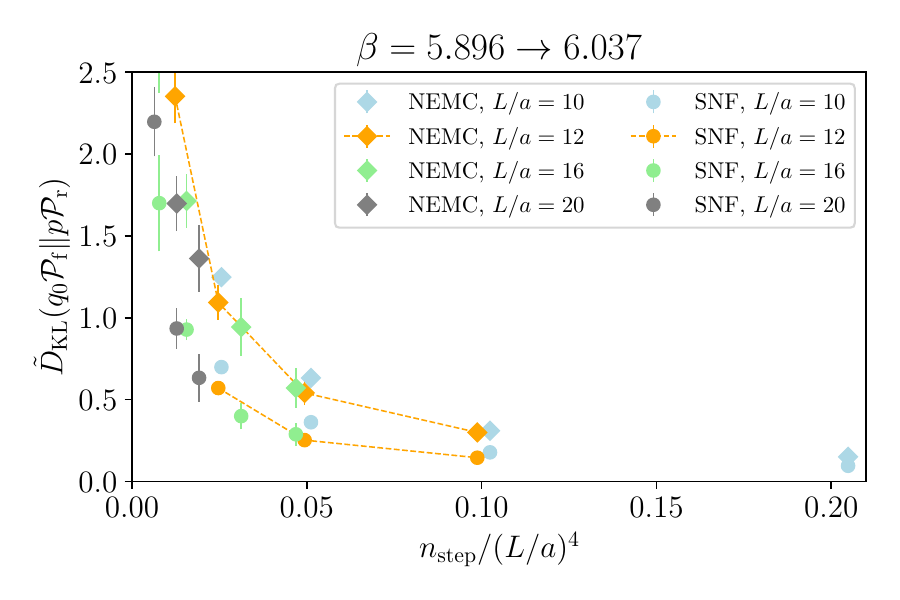}
 \includegraphics[scale=0.55,keepaspectratio=true]{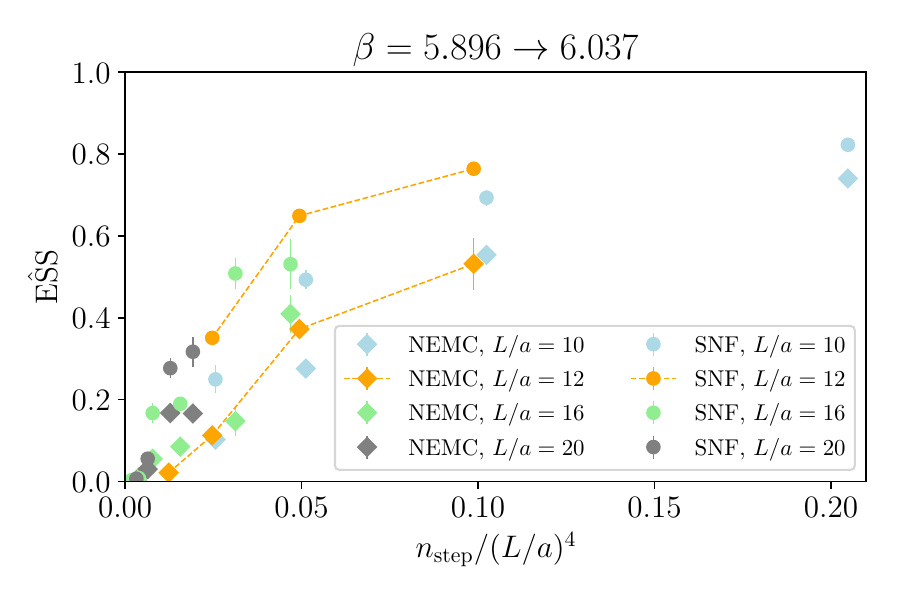}
 \caption{Results for the KL divergence (top panel) and for the ESS (bottom panel) for all the lattice sizes analyzed in this study for a flow between $\beta=5.896$ and $\beta=6.037$, plotted against $\nstep/(L/a)^4$, for NE-MCMC (diamonds) and SNF (circles) architectures. $L/a=12$ data are connected with a dashed line to guide the eye.}
 \label{fig:scaling_dkless_a2}
\end{figure}

We report in fig.~\ref{fig:scaling_dkless_a2} results for the scaling in the volume for the `a2' ensemble of out-of-equilibrium evolutions, which are conducted at coarser lattice spacings. No noticeable qualitative differences emerge with respect to the `a3' ensemble discussed in the main text.

\begin{figure}[t]
 \includegraphics[scale=0.55,keepaspectratio=true]{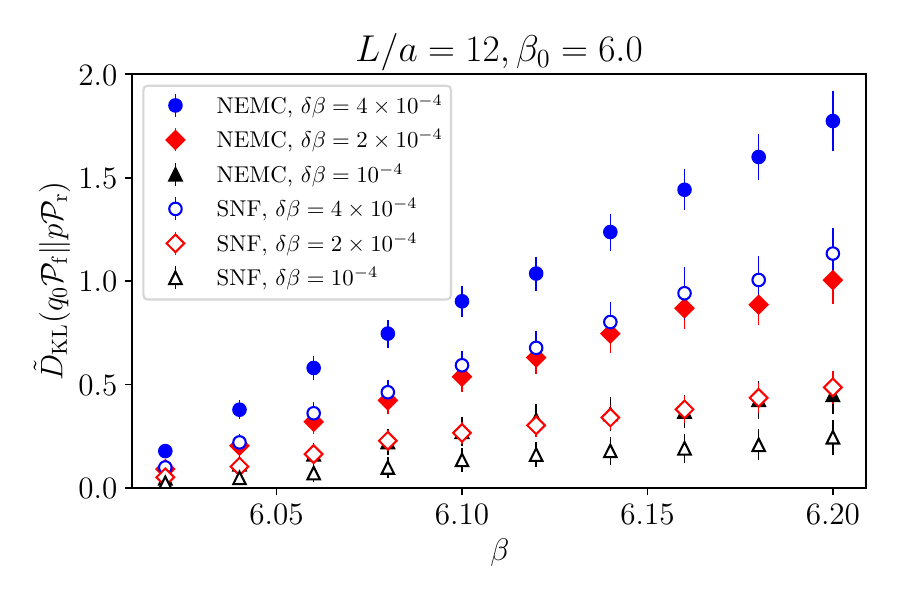}
 \includegraphics[scale=0.55,keepaspectratio=true]{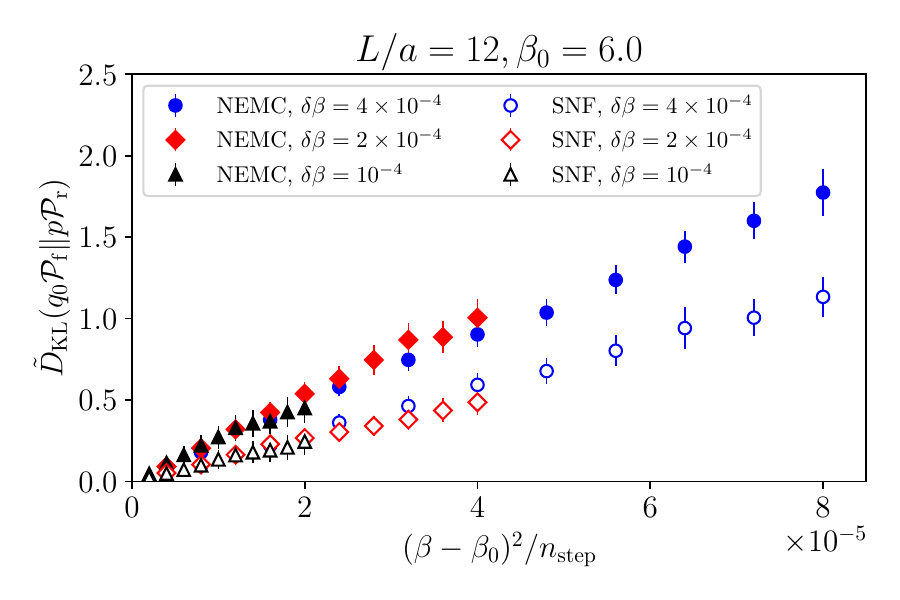}
 \caption{Results for the KL divergence for $L/a=12$ for flows starting from $\beta=6.0$, with linear protocol and different increases $\delta \beta = (\beta - \beta_0)/\nstep$ at each step. Results are plotted against the inverse coupling (top panel) or against the change in inverse coupling normalized by $\delta \beta$ (bottom panel).}
 \label{fig:scaling_beta}
\end{figure}

Furthermore, we complement our discussion with an analysis of the scaling of both NE-MCMC and SNFs in the inverse coupling $\beta$. On one hand this must be considered a preliminary analysis, as the behavior of these flow-based approaches may change when different, more efficient protocols are used. On the other hand, it is certainly useful to understand what happens when the target $\beta$ is further and further away from the starting probability distribution.
In fig.~\ref{fig:scaling_beta} the KL divergence is plotted as a function of the value of $\beta$ of the target distributions. These plots have been obtained using intermediate values of $\beta$ of transformations of the `b0' ensemble at fixed volume, see table~\ref{tab:ensembles}: we stress here that one of the advantages of both NE-MCMC and SNFs is that intermediate values of the parameter(s) changed throughout an evolution to a given target can be sampled as well.

It is interesting to note that, at least for a linear protocol and in this range of lattice spacings, the KL divergence increases roughly linearly with $\beta$: transformations with different increments $\delta \beta = (\beta_t - \beta_0)/\nstep$ are characterized by different slopes. These results suggest that in order to sample larger values of $\beta$ with these flows with a target value of the KL divergence, even smaller values of $\delta \beta$ would be required.
To quantify this effort, in the bottom panel of fig.~\ref{fig:scaling_beta} we plot the KL divergence against the square of the change in $\beta$ normalized by $\nstep$: the results both for NE-MCMC and SNFs collapse on two distinct lines, indicating that, at least for a linear protocol and in this range of lattice spacings, to keep the KL divergence constant the number of steps must be scaled with $(\beta-\beta_0)^2$.
A complete analysis of the scaling in the continuum limit for evolutions in $\beta$ would have to take into account two non-trivial factors: the value of $\beta_0$ where the evolutions start and the fact that all intermediate values of $\beta$ can be in principle sampled in the same evolution. 
Furthermore, a full quantitative analysis of the scaling in $\beta$ would necessarily include a thorough study of the protocol used to change the inverse coupling (or any other parameter).

\section{Work distributions and the free energy}
\label{sec:work_distro}

\begin{figure}[t]
 \includegraphics[scale=0.55,keepaspectratio=true]{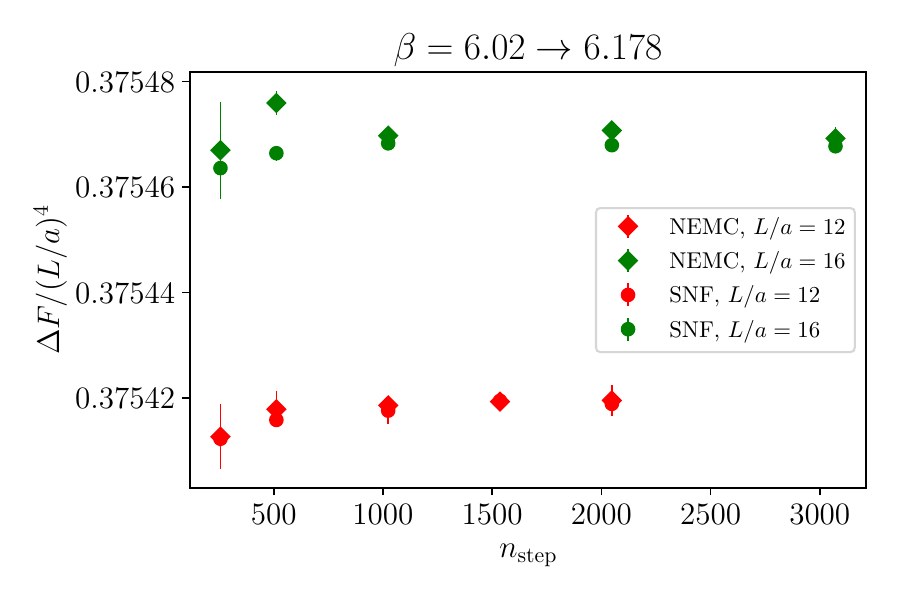}
 \caption{Results for the dimensionless free energy $\Delta F$  for a flow between $\beta=6.02$ and $\beta=6.178$, for $L/a=12$ and $L/a=16$, for NE-MCMC (diamonds) and SNF (circles) architectures.}
 \label{fig:deltaf_a3}
\end{figure}

Understanding the behavior of the average over non-equilibrium evolutions that we perform, for example, when applying Jarzynski's equality or the reweighting formula of eq.~\eqref{eq:estimator} is instrumental to obtain reliable results with this approach. The free-energy $\Delta F$ obtained directly from eq.~\eqref{eq:jar} is the most natural candidate, as it has to be the same irrespective of the architecture being used. Results reported in fig.~\ref{fig:deltaf_a3} have been obtained both with NE-MCMC and SNFs for a variable amount of $\nstep$ and show a general good agreement for both lattice sizes under study. Even if the computational effort is similar among the different architectures, flows with smaller $\nstep$ are less efficient and generally lead to larger errors. Furthermore, flows characterized by large $\DKL$ or small $\ESS$ generally are at risk of overestimating $\Delta F$ if the number of evolutions is too low: as it will be clear in the following, this exponential average is susceptible to the presence of ``rare events''.

\begin{figure}[t]
 \includegraphics[scale=0.55,keepaspectratio=true]{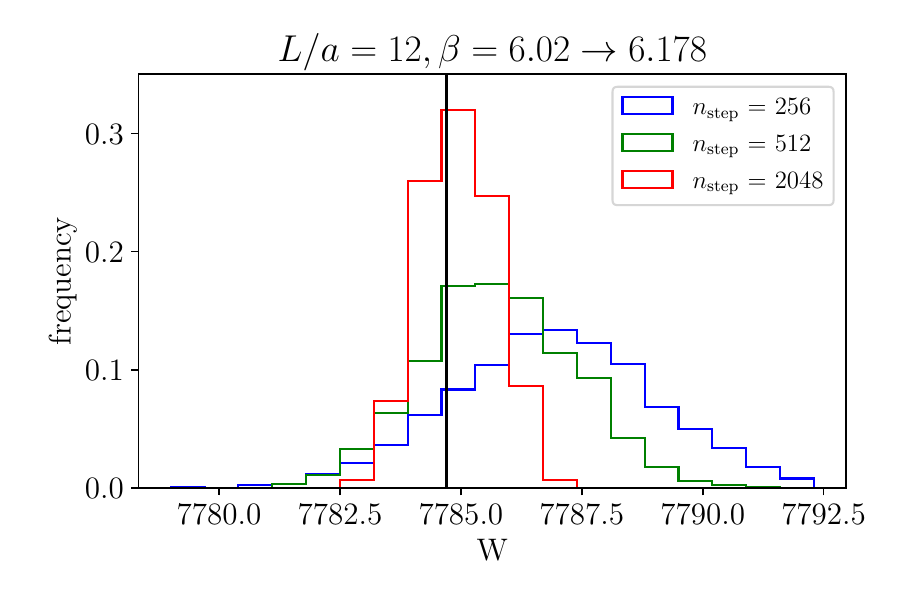}
 \includegraphics[scale=0.55,keepaspectratio=true]{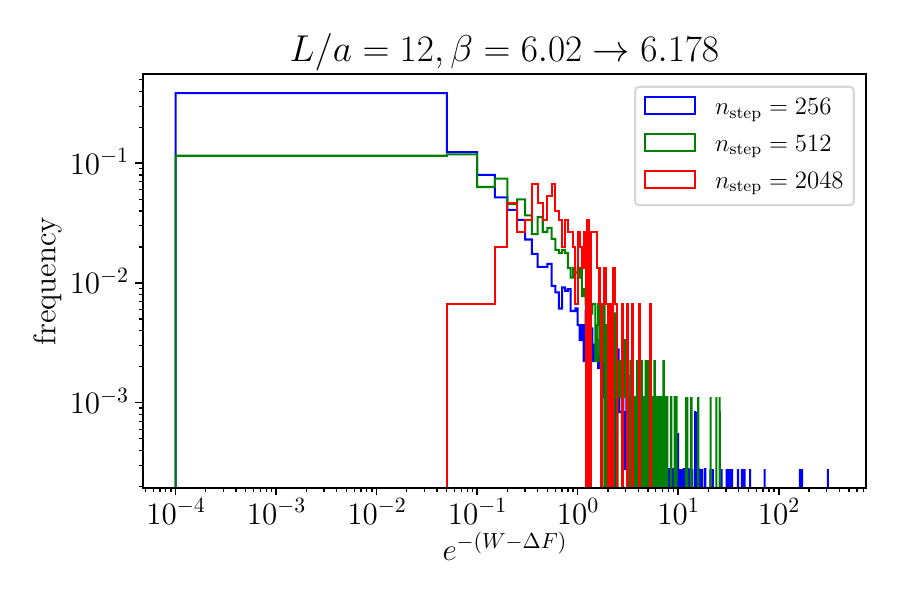}
 \caption{Distribution of the work $W$ (top panel) and of the exponential of the dissipated work $W - \Delta F$ (bottom panel) for $L/a=12$ for flows performed with NE-MCMC between $\beta=6.02$ and $\beta=6.178$ for 3 different values of $\nstep$. The black vertical line in the top panel represents the value of $\Delta F$ for this particular transformation in $\beta$, chosen to be the most precise value for NE-MCMC from fig.~\ref{fig:deltaf_a3}. The sum of the histogram bars equals unity in all cases.}
 \label{fig:histograms}
\end{figure}

From a qualitative perspective, it is very useful to analyze the distribution of the work $W$ for evolutions with different values of $\nstep$, that we present in fig.~\ref{fig:histograms} for the case of the NE-MCMC simulations. It is immediate to see that for evolutions closer to equilibrium the work has a smaller variance and, at the same time, the average is closer to $\Delta F$. This behavior is perfectly consistent with the approximate relation
\begin{equation}
    \langle W \rangle_{\rm f} - \Delta F \simeq \frac{1}{2} \Var(W);
\end{equation}
we refer to the analysis of ref.~\cite{Bonanno:2024udh} for an example of a quantitative analysis of this relation. 

Furthermore, the distribution of $\exp(-W)$ is also crucial, as it provides an immediate understanding of the quality of the sampling using eq.~\eqref{eq:estimator}. In particular, in the bottom panel of fig.~\ref{fig:histograms} we show how the values of $\exp(-W_{\rm d})$ from different evolutions behave for various values of $\nstep$ for a fixed transformation. While being further away for equilibrium (i.e., for smaller $\nstep$) the distribution is highly asymmetric. This means that the estimation of $\Delta F$ with eq.~\eqref{eq:jar} and the sampling with eq.~\eqref{eq:estimator} are reliable only if the tail on the right is populated enough, i.e., there are enough evolutions with $W < \Delta F$ (the ``rare events'' mentioned at the beginning of this section). On the other hand, when the transformation is closer to a reversible one (i.e., for the largest value of $\nstep$) the variance of the distribution is much smaller and less evolutions are needed for a correct estimation of observables and of the ratio of partition functions. This behavior is consistently described by the $\hat\ESS$, which precisely monitors the variance of this distribution, as understood in eq.~\eqref{eq:essvar}.

\begin{acknowledgments}
We thank M.~Albergo, S.~Bacchio, C.~Bonanno, M.~Caselle, L.~Giusti, G.~Kanwar, P.~Kessel, S.~Nakajima, K.~Nicoli, M.~Panero, D.~Panfalone, A.~Patella, S.~Schaefer, A.~Tomiya, D.~Vadacchino, L.~Vaitl and L.~Verzichelli for insightful comments and discussions. The numerical simulations were run on machines of the Consorzio Interuniversitario per il Calcolo Automatico dell'Italia Nord Orientale (CINECA). E.~Cellini and A.~Nada acknowledge support and A.~Bulgarelli acknowledges partial support by the Simons Foundation grant 994300 (Simons Collaboration on Confinement and QCD Strings).  A.~Nada acknowledges support from the European Union - Next Generation EU, Mission 4 Component 1, CUP D53D23002970006, under the Italian PRIN “Progetti di Ricerca di Rilevante Interesse Nazionale – Bando 2022” prot. 2022ZTPK4E. All the authors acknowledge support from the SFT Scientific Initiative of INFN. 
\end{acknowledgments}

\bibliography{biblio}

\end{document}